\documentclass[12pt]{amsart}
\usepackage{amssymb}

\theoremstyle{plain}

\numberwithin{equation}{section}

\begin{document}
\title{Polyanalytic relativistic second Bargmann transforms}
\author{Zouha\"{i}r Mouayn}
\address{{\footnotesize{Faculty of Sciences and Technics (M'Ghila), PO. Box. 523, B\'{e}ni
Mellal, Morocco.}}}

\maketitle

\begin{abstract}
We construct coherent states through special superpositions of photon number
states of the relativistic isotonic oscillator. In each superposition the
coefficients are chosen to be $L^{2}$-eingenfunctions of a $\sigma $-weight
Maass Laplacian on the Poincar\'{e} disk, which are associated with the
eigenvalue $4m\left( \sigma -1+m\right) $, $m\in \mathbb{Z}_{+}\cap \left[
0,\left( \sigma -1\right) /2\right] $. For each nonzero $m$ the associated
coherent states transform constitutes the $m$-\textit{true}-polyanalytic
extension of a relativistic version of the second Bargmann transform, whose
integral kernel is expressed in terms of a special Appel-Kamp\'{e} de
F\'{e}riet's hypergeometric function. The obtained results could be used to
extend the known semi-classical analysis of quantum dynamics of the
relativistic isotonic oscillator.
\end{abstract}

\section{Introduction}

The second Bargmann transform $\left[ 1\right] $ can be introduced in a
slight modified form $\left[ 2\right] $ as
\begin{equation}
\mathcal{B}_{\sigma }:L^{2}\left( \mathbb{R}_{+},d\rho \right) \rightarrow
\mathcal{A}^{\sigma }\left( \mathbb{D}\right)   \tag{1.1}
\end{equation}
defined by
\begin{equation}
\mathcal{B}_{\sigma }\left[ \phi \right] \left( z\right) :=\sqrt{\frac{%
\sigma -1}{\pi \Gamma \left( \sigma \right) }}\left( 1-z\right) ^{-\sigma
}\int\limits_{0}^{+\infty }\exp \left( -\frac{1}{2}\rho \left( \frac{1+z}{1-z%
}\right) \right) \phi \left( \rho \right) \rho ^{\frac{1}{2}(\sigma
-1)}d\rho   \tag{1.2}
\end{equation}
where $\sigma >1$ is a fixed parameter,
\begin{equation}
\mathcal{A}^{\sigma }\left( \mathbb{D}\right) :=\left\{ f\text{ analytic on }%
\mathbb{D}\text{, }\int_{\mathbb{D}}\left| f\left( z\right) \right|
^{2}\left( 1-z\overline{z}\right) ^{\sigma -2}d\mu \left( z\right) <+\infty
\right\}   \tag{1.3}
\end{equation}
denotes the weighted Bergman space on the unit disk $\mathbb{D}=\left\{ z\in
\mathbb{C};\left| z\right| <1\right\} $ and $d\mu $ being the Lebesgue
measure on it. The involved kernel function in $\left( 1.2\right) $
corresponds to the generating function of Laguerre polynomials $\left[ 3%
\right] $ which turn out to be fundamental pieces in expressing eigenstates
of the isotonic oscillator that is the harmonic oscillator with an inverse
quadratic non-linear term (like a centripetal barrier) $\left[ 4\right] $.
Now, after noticing that the Laguerre polynomials can also be obtained as a
special limit of continuous dual Hahn polynomials $\left[ 5\right] $ which
are involved in eigenstates of the relativistic isotonic oscillator $\left[ 6%
\right] $, we here propose a \textit{true} \textit{polyanalytic} extension
of order $m$ for a \textit{relativistic} version of the second Bargmann
transform by replacing the arrival space in $\left( 1.1\right) $ by the
eigenspace $\left[ 7\right] :$%
\begin{equation}
\mathcal{A}_{m}^{\sigma }\left( \mathbb{D}\right) :=\left\{ f:\mathbb{D}%
\rightarrow \mathbb{C}\text{, }\Delta _{\sigma }f=\epsilon _{m}f\text{, }%
\int_{\mathbb{D}}\left| f\left( z\right) \right| ^{2}\left( 1-z\overline{z}%
\right) ^{\sigma -2}d\mu \left( z\right) <+\infty \right\}   \tag{1.4}
\end{equation}
of the $\sigma $-weight Maass Laplacian
\begin{equation}
\Delta _{\sigma }:=-4\left( 1-z\overline{z}\right) \left( \left( 1-z%
\overline{z}\right) \frac{\partial ^{2}}{\partial z\partial \overline{z}}%
-\sigma \overline{z}\frac{\partial }{\partial \overline{z}}\right)
\tag{1.5}
\end{equation}
with the eigenvalue
\begin{equation}
\epsilon _{m}:=4m\left( \sigma -1-m\right) ,\text{ \ }m=0,1,2,...,\left%
\lfloor \left( \sigma -1\right) /2\right\rfloor,  \tag{1.6}
\end{equation}
where $\left\lfloor .\right\rfloor $ is the greatest integer function. \ The
operator $\Delta _{\sigma }$ can be unitarily intertwined to represent the
Schr\"{o}dinger operator of a charged particle evolving in the Poincar\'{e}
disk under the action of a constant homogeneous magnetic field with a
strength proportional to\ $\sigma $.

For $m=0,$ the space $\mathcal{A}_{0}^{\sigma }\left( \mathbb{D}\right) $ in
$\left( 1.4\right) $ coincides with the Bergman space $\mathcal{A}^{\sigma
}\left( \mathbb{D}\right) $ in $\left( 1.3\right) $ and for a nonzero $m$
the space $\mathcal{A}_{m}^{\sigma }\left( \mathbb{D}\right) $ is the true
polyanalytic Bergman space of order $m$ (see $\left[ 8\right] $). We
precisely construct a family of integral transforms with the form
\begin{equation*}
\mathcal{B}_{rel,\nu }^{m}:L^{2}\left( \mathbb{R}_{+},d\rho \right)
\rightarrow \mathcal{A}_{m}^{\sigma }\left( \mathbb{D}\right)
\end{equation*}
defined by
\begin{equation}
\mathcal{B}_{rel,\nu }^{m}\left[ \phi \right] \left( z\right) =\frac{\left(
\frac{2\nu -1}{\pi }\right) ^{\frac{1}{2}}\sqrt{2\Gamma \left( 2\nu
+m\right) }}{\Gamma \left( \nu +\frac{1}{2}\right) \Gamma \left( 2\nu
\right) \sqrt{m!}}\left( \frac{1}{1-z}\right) ^{2\nu }\left( \frac{\overline{%
z}-1}{\left( 1-z\right) \left( 1-z\overline{z}\right) }\right) ^{m}
\tag{1.7}
\end{equation}
\begin{equation*}
\times \int\limits_{0}^{+\infty }\digamma _{5}\left(
\begin{array}{ccc}
\nu -i\rho & \nu +i\rho : & 2\nu +m \\
. & \nu +\frac{1}{2}: & 2\nu
\end{array}
\mid \frac{1-\overline{z}z}{\left( 1-\overline{z}\right) \left( z-1\right) },%
\frac{1}{1-\overline{z}}\right) \left( c^{2}\right) ^{i\rho }\frac{\Gamma
^{2}\left( \nu -i\rho \right) }{i^{\nu }\Gamma \left( -i\rho \right) }\phi
\left( \rho \right) d\rho
\end{equation*}
for any $z\in \mathbb{D}$. Here, $\sigma =2\left( \nu +m\right) $, $\nu =(1+%
\sqrt{1+2c^{4}})/2$, $c>0$ is the light velocity and $\digamma _{5}$\textit{%
\ }is a special Appel-Kamp\'{e} de F\'{e}riet hypergeometric function $\left[
9\right] $.\medskip

The construction of the integral transform $\left( 1.7\right) $ is based on
a coherent states analysis by adopting a general Hilbertian probabilistic
scheme $\left[ 10\right] $. That is, we define a coherent state through a
special superposition of photon number states of the relativistic isotonic
oscillator by choosing the coefficients to be $L^{2}$-eigenfunctions of $%
\Delta _{\sigma }$ associated with the eigenvalue $\epsilon _{m}$ in $\left(
1.6\right) $ and forming an orthonormal basis of the space $\mathcal{A}%
_{m}^{\sigma }\left( \mathbb{D}\right) $. Actually, for a nonzero integer $m$
the associated coherent states transform $\left( 1.7\right) $ could be
considered as a true polyanalytic extension for the relativistic version,
say $\mathcal{B}_{rel,\nu }^{0}$, of the second Bargmann transform $\left(
1.1\right) $. \

We should also note that in the analytic case $m=0$, the $\digamma _{5}$-sum
in $\left( 1.7\right) $ reduces to a Gauss hypergeometric $_{2}\digamma _{1}$%
-sum $\left[ 5\right] $ making possible to assert that the transform $%
\mathcal{B}_{rel,\nu }^{0}$ can also be derived from the $SU(1,1)$ coherent
states introduced by the authors $\left[ 11\right] $ who also were concerned
by evaluating some path integrals involved in the semiclassical dynamics of
the relativistic isotonic oscillator. This also means that the generalized
coherent states with their transforms we here are introducing could be
exploited as news basic tools in order to extend the description of the
known semi-classical portraits of quantum dynamics for the relativistic
isotonic oscillator.\medskip \newpage

The paper is organized as follows. In Section 2, we recall briefly some
needed tools from the $L^{2}$-spectral theory of the Maass Laplacian on the
Poincar\'{e} disk. To each fixed discrete eigenvalue of this operator a set
of generalized $SU\left( 1,1\right) $ coherent states will be attached in
section 3 without specifying the Hamiltonian system. \ In section 4 we
review the definition of the relativistic isotonic oscillator as well as
some of its basic ingredients. For this oscillator, we construct in section
5 a set of generalized coherent states and we obtain expressions for their
wave functions leading to define the corresponding coherent states
transforms.

\section{Maass Laplacians on the Poincar\'{e} disk}

The unit disk $\mathbb{D}$ is endowed with its usual Kh\"{a}ler metric $%
ds^{2}=-\partial \overline{\partial }\mathrm{Log}\left( 1-z\overline{z}%
\right) dz\otimes d\overline{z},$ the hyperbolic distance on $\mathbb{D}$ is
given by
\begin{equation}
\cosh ^{2}d\left( z,w\right) =\frac{(1-z\overline{w})(1-\overline{z}w)}{%
\left( 1-z\overline{z}\right) \left( 1-w\overline{w}\right) }  \tag{2.1}
\end{equation}
and the volume element reads
\begin{equation}
d\mu _{0}\left( z\right) :=\left( 1-z\overline{z}\right) ^{-2}d\mu \left(
z\right) \text{.}  \tag{2.2}
\end{equation}
Let us consider the $1-$form on $\mathbb{D}$ defined by $\theta :=-i\left(
\partial -\overline{\partial }\right) \mathrm{Log}\left( 1-z\overline{z}%
\right) $ to which the Schr\"{o}dinger operator
\begin{equation}
H_{\sigma }:=\left( d+\frac{i}{2}\sigma ext\left( \theta \right) \right)
^{\ast }\left( d+\frac{i}{2}\sigma ext\left( \theta \right) \right)
\tag{2.3}
\end{equation}
can be associated. Here $d$ denotes the
usual exterior derivative on differential forms on $\mathbb{D}$ and $%
ext\left( \theta \right) $ is the exterior multiplication by $\theta $ while
the symbol $\ast $ stands for the adjoint operator with respect to the
Hermitian scalar product induced by the metric $ds^{2}$ on differential
forms. Actually, the operator $H_{\sigma }$ is acting on the Hilbert space $%
L^{2}\left( \mathbb{D},d\mu _{0}\left( z\right) \right) $ and can be
unitarily intertwined as
\begin{equation}
\left( 1-z\overline{z}\right) ^{\frac{1}{2}\sigma }\Delta _{\sigma }\left(
1-z\overline{z}\right) ^{-\frac{1}{2}\sigma }=H_{\sigma }  \tag{2.4}
\end{equation}
in terms of the Maass Laplacian $\Delta _{\sigma }$ in $\left( 1.5\right) $.
The latter is acting on the Hilbert space $L^{2,\sigma }\left( \mathbb{D}%
\right) $ of square integrable functions on $\mathbb{D}$ with respect to the
measure
\begin{equation}
d\mu _{\sigma }\left( z\right) :=\left( 1-z\overline{z}\right) ^{\sigma
-2}d\mu \left( z\right).  \tag{2.5}
\end{equation}
Different aspects of the spectral analysis of $\Delta _{\sigma }$ have been
studied by many authors (see $\left[ 2\right] $, $\left[ 12\right] $ and
references therein). Note that $\Delta _{\sigma }$ is an elliptic densely
defined operator on $L^{2,\sigma }\left( \mathbb{D}\right) $ and admits a
unique self-adjoint realization that we denote also by $\Delta _{\sigma }$.
Its spectrum consists of a continuous part $\left[ 1,+\infty \right[ $
corresponding to \textit{scattering states} and a finite number of
eigenvalues $\left\{ \epsilon _{m}\right\} $ in $\left( 1.6\right) $
occurring with infinite degeneracy, provided that $\sigma >1$. To these
eigenvalues correspond \textit{bound states} belonging to the eigenspace $%
\left( 1.4\right) $ which simply reads
\begin{equation}
\mathcal{A}_{m}^{\sigma }(\mathbb{D}):=\left\{ f\in L^{2,\sigma }\left(
\mathbb{D}\right) \text{ and }\Delta _{\sigma }f=\epsilon _{m}f\right\}
\tag{2.6}
\end{equation}
Elements of an orthogonal basis of this space are of the form
\begin{equation}
l_{k}^{(\sigma ,m)}\left( z\right) :=\frac{\left( -1\right) ^{\min \left(
m,k\right) }}{\left( 1-z\overline{z}\right) ^{m}}\left| z\right| ^{\left|
m-k\right| }e^{-i\left( m-k\right) \arg z}P_{\min \left( m,k\right)
}^{\left( \left| m-k\right| ,\sigma -2m-1\right) }\left( 1-2z\overline{z}%
\right) \text{, \ }k=0,1,...,  \tag{2.7}
\end{equation}
\ where $P_{n}^{\left( \tau ,\beta \right) }\left( .\right) $ are Jacobi
polynomials $\left[ 3\right] $. Their norm square in $L^{2,\sigma }\left(
\mathbb{D}\right) $ is given by
\begin{equation}
\left\| l_{k}^{(\sigma ,m)}\right\| ^{2}=\pi \left( \sigma -2m-1\right) ^{-1}%
\frac{\left( \max (m,k\right) )!\Gamma \left( \sigma -2m+\min (m,k\right) )}{%
\left( \min (m,k\right) )!\Gamma \left( \sigma -2m+\max (m,k\right) )}.
\tag{2.8}
\end{equation}
Thus, the set of functions
\begin{equation}
f_{k}^{(\sigma ,m)}:=\frac{l_{k}^{(\sigma ,m)}}{\left\| l_{k}^{(\sigma
,m)}\right\| }\text{, \ \ \ }k=0,1,2,...\text{ ,}  \tag{2.9}
\end{equation}
is an orthonormal basis of $\mathcal{A}_{m}^{\sigma }\left( \mathbb{D}%
\right) $ and can be rewritten explicitly as
\begin{equation}
f_{k}^{(\sigma ,m)}\left( z\right) =\left( -1\right) ^{k}\sqrt{\frac{\left(
\sigma -2m-1\right) \Gamma \left( \sigma -m\right) k!}{\pi m!\Gamma \left(
\sigma -2m+k\right) }}\frac{\overline{z}^{m-k}}{\left( 1-z\overline{z}%
\right) ^{m}}P_{k}^{\left( m-k,\sigma -2m-1\right) }\left( 1-2z\overline{z}%
\right)  \tag{2.10}
\end{equation}
by making use of the connection formula ($\left[ 13\right] $, p.556):
\begin{equation}
P_{m}^{\left( k-m,\beta \right) }\left( u\right) =\frac{k!\Gamma \left(
m+\beta +1\right) }{m!\Gamma \left( k+\beta +1\right) }\left( \frac{u-1}{2}%
\right) ^{m-k}P_{k}^{\left( m-k,\beta \right) }\left( u\right)  \tag{2.11}
\end{equation}
for $u=1-2z\overline{z}$ and $\beta =2\left( \nu -m\right) -1$.

Note that when $m=0$, the basis elements $\left( 2.10\right) $ reduce to
\begin{equation}
f_{k}^{(\sigma ,0)}\left( z\right) =\left( \sigma -1\right) ^{\frac{1}{2}}%
\sqrt{\frac{\Gamma \left( \sigma +k\right) }{\pi k!\Gamma \left( \sigma
\right) }}z^{k}  \tag{2.12}
\end{equation}
and constitute a basis of the Bergman space $L_{a}^{2,\sigma }\left( \mathbb{%
D}\right) $ of analytic square integrable functions on $\mathbb{D}
$ with respect to the measure $d\mu _{\sigma }$ in (2.5). This means that $%
L_{a}^{2,\sigma }\left( \mathbb{D}\right) \equiv $ $\mathcal{A}_{0}^{\sigma
}\left( \mathbb{D}\right) $. For $m\neq0$ the spaces $\mathcal{A}%
_{m}^{\sigma }\left( \mathbb{D}\right) $ are called polyanalytic Bergman
spaces and more precisely, they are the true polyanalytic ones (see $\left[ 8%
\right] $, pp.32-35 and references therein).\medskip

\textbf{Remark 2.1}. The operator $\Delta _{\sigma }$ in $\left( 1.5\right) $
can also be obtained $\left[ 12\right] $ from the $\sigma $-weight Maass
Laplacian $y^{2}\left( \partial _{x}^{2}+\partial _{y}^{2}\right) -i\sigma
y\partial _{x}$ on the Poincar\'{e} upper half-plane $\left[ 14\right] $.
The condition $\sigma >1$ ensuring the existence of eigenvalues $\epsilon
_{m}$ (\textit{hyperbolic Landau levels}) should implies that the magnetic
field $B=\sigma \Omega \left( z\right) $, where $\Omega $ stands for the
Kh\"{a}ler 2-form on $\mathbb{D}$, has to be strong enough to capture the
particle in a closed orbit. If this condition is not fulfilled the motion
will be unbounded and the classical orbit of the particle will intercept the
disk boundary whose points stands for $\left\{ \infty \right\} $ which means
escaping to infinity $\left[ 15\right] $.

\section{Generalized $SU(1,1)$ coherent states}

The $SU(1,1)$ discrete series coherent states or negative binomial states
(NBS) $\left[ 16\right] $ are labelled by points $z$ $\in \mathbb{D}$ and
are an example of nonlinear coherent states $\left[ 17\right] $ with the
form
\begin{equation}
\mid z,\sigma >:=\left( 1-z\overline{z}\right) ^{\frac{1}{2}\sigma
}\sum\limits_{k=0}^{+\infty }\sqrt{\frac{\Gamma \left( \sigma +k\right) }{%
\Gamma \left( \sigma \right) k!}}z^{k}\mid k>  \tag{3.1}
\end{equation}
where $\sigma >1$ is a fixed parameter and the \textit{kets} $\left\{ \mid
k>\right\} $ denote for instance Fock states in a Hilbert space $%
\mathcal{H}$. Observe that the coefficients in the superposition $\left(
3.1\right) :$
\begin{equation}
f_{k}^{(\sigma ,0)}\left( z\right) \varpropto \sqrt{\frac{\Gamma \left(
\sigma +k\right) }{\pi \Gamma \left( \sigma \right) k!}}z^{k}\text{, \ \ \ \
\ }k=0,1,2,...  \tag{3.3}
\end{equation}
constitute a basis of the weighted Bergman space $L_{a}^{2,\sigma }\left(
\mathbb{D}\right) $ analytic functions belonging to $L^{2,\sigma
}\left( \mathbb{D}\right) $. This space was associated with the first
eigenvalue $\epsilon _{0}=0$ corresponding to $m=0$. This observation
suggests to replace the coefficients $\left( 3.3\right) $ by basis elements $%
f_{k}^{\sigma ,m}\left( z\right) $ of the polyanalytic Bergman space\ $%
\mathcal{A}_{m}^{\sigma }\left( \mathbb{D}\right) $ allowing us to consider
a generalization of the states $\left( 3.1\right) $ through the following
superposition
\begin{equation}
\mid z,\sigma ,m>=\left( \mathcal{N}_{\sigma ,m}\left( z\right) \right) ^{-%
\frac{1}{2}}\sum\limits_{k=0}^{+\infty }\overline{f_{k}^{(\sigma ,m)}\left(
z\right) }\mid k>  \tag{3.4}
\end{equation}
where
\begin{equation}
\mathcal{N}_{\sigma ,m}\left( z\right) =\pi ^{-1}(\sigma -2m-1)\left( 1-z%
\overline{z}\right) ^{-\sigma }  \tag{3.5}
\end{equation}
is a normalization factor. Here, the states in $\left( 3.4\right) $ will be
indicated by GNBS for brevity. The overlap relation between two GNBS is
given by
\begin{equation}
<w,\sigma ,m\mid z\text{ },\sigma ,m>_{\mathcal{H}}=\frac{\Gamma \left(
\sigma -m\right) \left( \left( 1-z\overline{z}\right) \left( 1-w\overline{w}%
\right) \right) ^{\frac{1}{2}\sigma }}{\left( -1\right) ^{m}m!\Gamma \left(
\sigma -2m\right) \left( 1-z\overline{w}\right) ^{\sigma }}  \tag{3.6}
\end{equation}
\begin{equation*}
\times \left( \frac{(1-z\overline{w})\left( 1-\overline{w}z\right) }{\left(
1-z\overline{z}\right) \left( 1-w\overline{w}\right) }\right) ^{m}\
_{2}\digamma _{1}\left( -m,\sigma -m,\sigma -2m;\frac{\left( 1-z\overline{z}%
\right) \left( 1-w\overline{w}\right) }{(1-z\overline{w})\left( 1-\overline{w%
}z\right) }\right)
\end{equation*}
where $_{2}\digamma _{1}$\ is a terminating Gauss hypergeometric sum $\left[
5\right] $. Actually, the basic minimum properties for these states to be
considered as coherent states are satisfied $\left[ 7\right] $. Namely, the
conditions which have been formulated by Klauder $\left[ 18\right] $: $%
\left( a\right) $ the continuity of labelling, $\left( b\right) $ the fact
that these states are normalizable but not orthogonal and $\left( c\right) $
these states fulfilled the resolution of the identity as
\begin{equation}
\mathbf{1}_{\mathcal{H}}=\pi ^{-1}(\sigma -2m-1)\int\limits_{\mathbb{D}}\mid
z,\sigma ,m><z,\sigma ,m\mid d\mu _{0}\left( z\right)  \tag{3.7}
\end{equation}
in terms of the measure $d\mu _{0}$\textit{\ }given in $\left( 2.2\right) $.
\textit{\ }

As we can see, these coherent states are independent of the basis $\left\{
\mid k>\right\} $ we use and the only condition which is implicitly
fulfilled is orthonormality relations of its elements in $\mathcal{H}$. But
if we want to attach to these GNBS a concrete quantum system then a
Hamiltonian operator should be specified together with a corresponding
explicit eigenstates basis. As example, this was done in $\left[ 7\right] $
for the isotonic oscillator
\begin{equation}
H_{\tau }:=-\frac{1}{2}\partial _{x}^{2}+\frac{\tau ^{2}-1/4}{2x^{2}}+\frac{1%
}{2}x^{2}\text{, \ \ \ \ }\tau \geq 1/2,  \tag{3.8}
\end{equation}
with the associated states Hilbert space\ $L^{2}\left( \mathbb{R}%
_{+},dx\right) $. \ Here our next task is to particularize the formalism of
the above GNBS for a\textit{\ relativistic} version of the isotonic
oscillator $\left( 3.8\right) $.

\section{A relativistic isotonic oscillator}

In this section, we briefly recall the definition and eigenstates of the
relativistic isotonic oscillator as presented in $\left[ 6\right] $ where it
was called the relativistic linear singular oscillator and given by the
following finite-difference operator
\begin{equation}
H:=m_{\ast }c^{2}\left( \cosh i\partial _{\rho }+\frac{1}{2}\omega
_{0}^{2}\rho ^{\left( 2\right) }e^{i\partial _{\rho }}+\frac{g_{0}}{\rho
^{\left( 2\right) }}e^{i\partial _{\rho }}\right)   \tag{4.1}
\end{equation}
in terms of the dimensionless variable $\rho =x/\lambda $ where $\lambda
=\hbar /m_{\ast }c$ is the Compton wavelength of the particle and parameters
$\omega _{0}=\hbar \omega /m_{\ast }c^{2}$, $g_{0}=m_{\ast }g/\hbar ^{2}$.
Here, $\rho ^{\left( \delta \right) }$ denotes the generalized degree ($%
\left[ 19\right] $, p.201):
\begin{equation}
\rho ^{\left( \delta \right) }:=i^{\delta }\frac{\Gamma \left( \delta -i\rho
\right) }{\Gamma \left( -i\rho \right) }  \tag{4.2}
\end{equation}
for $\delta =2.$ The eigenfunctions of $H$ obeying the Dirichlet boundary
conditions on the interval $\left[ 0,\infty \right) $ are given by
\begin{equation}
\psi _{k}\left( \rho \right) =c_{k}\omega _{0}^{i\rho }\left( -\rho \right)
^{\left( \alpha \right) }\Gamma \left( \nu +i\rho \right) S_{k}\left( \rho
^{2},\alpha ,\nu ,\frac{1}{2}\right)   \tag{4.3}
\end{equation}
where $c_{k}$ is a constant given by
\begin{equation}
c_{k}=\sqrt{2}\left( k!\Gamma \left( k+\alpha +\nu \right) \Gamma \left(
k+\alpha +1/2\right) \Gamma \left( k+\nu +1/2\right) \right) ^{-\frac{1}{2}}
\tag{4.4}
\end{equation}
and $S_{k}\left( t^{2},a,b,c\right) $ are the continuous dual Hahn
polynomials defined by ($\left[ 6\right] $, p.331):
\begin{equation}
S_{k}\left( t^{2},a,b,c\right) :=\left( a+b\right) _{k}\left( a+c\right)
_{k}{}\ _{3}\digamma _{2}\left(
\begin{array}{c}
-k,a+it,a-it \\
a+b,a+c
\end{array}
;1\right)   \tag{4.5}
\end{equation}
in terms of a terminating $_{3}\digamma _{2}$-sum. The eigenstates $\left\{
\psi _{k}\right\} $ are associated with eigenvalues
\begin{equation}
E_{k}:=\hbar \omega (2k+\alpha +\nu )\text{, }\ k=0,1,2,...  \tag{4.6}
\end{equation}
where
\begin{equation}
\alpha =\frac{1}{2}+\frac{1}{2}\sqrt{1+\frac{2}{\omega _{0}^{2}}\left( 1-%
\sqrt{1-8g_{0}\omega _{0}^{2}}\right) }\text{ and }\nu =\frac{1}{2}+\frac{1}{%
2}\sqrt{1+\frac{2}{\omega _{0}^{2}}\left( 1+\sqrt{1-8g_{0}\omega _{0}^{2}}%
\right) }.  \tag{4.7}
\end{equation}
By ($\left[ 20\right] $, p.317) the hermeticity condition imposes a
restriction on the values of the quantity $g_{0}$. Precisely, the
eigenvalues $E_{k}$ are real only in case when $\alpha $ and $\nu $ are real
or complex-conjugate, which imposes $g_{0}\geq 1/8\omega _{0}^{2}$. \

For our purpose we restrict ourselves to the lower bound case $%
g_{0}=1/8\omega _{0}^{2}$ and we choose the unit system $\hbar =m_{\ast
}=\omega =1$ in $\omega _{0}=\hbar \omega /m_{\ast }c^{2}.$ This means that
parameters $\left( 4.7\right) $ are now real with the form
\begin{equation}
\alpha =\nu \equiv \nu _{c}=\frac{1}{2}\left( 1+\sqrt{1+2c^{4}}\right) .
\tag{4.8}
\end{equation}
Under the above conditions we will be concerned with eigenstates $%
\left( 4.3\right) $ with the form
\begin{equation}
\psi _{k}^{\nu }\left( \rho \right) :=\frac{\sqrt{2}}{\Gamma \left( k+\nu
+1/2\right) \sqrt{k!\Gamma \left( k+2\nu \right) }}\left( c^{-2}\right)
^{i\rho }\left( -\rho \right) ^{\left( \nu \right) }\Gamma \left( \nu +i\rho
\right) S_{k}\left( \rho ^{2},\nu ,\nu ,\frac{1}{2}\right)   \tag{4.9}
\end{equation}
satisfying the orthonormality relations
\begin{equation}
\int\limits_{0}^{+\infty }\psi _{k}^{\nu }\left( \rho \right) \overline{\psi
_{j}^{\nu }\left( \rho \right) }d\rho =\delta _{k,j}  \tag{4.10}
\end{equation}
in the Hilbert space $L^{2}\left( \mathbb{R}_{+},d\rho \right) $.
Below, we will use the abreviation RIO for the relativistic isotonic
oscillator.\\
\textbf{Remark 4.1.} We should note that the eigenstates $\left\{ \psi
_{k}^{\nu }\right\} $ in $\left( 4.3\right) $ can also be obtained by the $k$%
-fold action of a finite-difference raising operator to the ground state,
see $\left[ 21\right] $ where the authors have established an exact
factorization of the RIO in a complete analogy with the non-relativistic
problem. In particular, they concluded that eigenfunctions $\left\{ \psi
_{k}^{\nu }\right\} $ constitute the basis of the irreducible representation
$D^{+}\left( \frac{\nu +\alpha }{2}\right) $ of the $SU(1,1)$ Lie group.

\section{Generalized $SU(1,1)$ coherent states for the RIO}

Here, we will proceed to attach the above generalized coherent states to the
RIO as follows.\medskip

\textbf{Definition\ 5.1. }\textit{For }$\sigma >1$\textit{\ and\ }$%
m=0,1,...,\left\lfloor \left( \sigma -1\right) /2\right\rfloor $.\textit{\ \
A class of coherent states can be defined through the superpositions}
\begin{equation}
\mid z;\left( \sigma ,m\right) ,\nu >=\left( \mathcal{N}_{\sigma ,m}\left(
z\right) \right) ^{-\frac{1}{2}}\sum_{k=0}^{+\infty }\overline{f_{k}^{\left(
\sigma ,m\right) }\left( z\right) }\mid \psi _{k}^{\nu }>  \tag{5.1}
\end{equation}
\textit{of the above eigenstates }$\mid \psi _{k}^{\nu }>$ \textit{of the RIO%
}, \textit{where }$\mathcal{N}_{\sigma ,m}\left( z\right) $\textit{\ is the
normalization factor in (3.5)} \textit{and} $f_{k}^{(\sigma ,m)}\left(
z\right) $\textit{\ are the basis elements of the space }$\mathcal{A}%
_{m}^{\sigma }(\mathbb{D})$, \textit{as defined by (2.9).\medskip }

We now assume that parameters $\sigma ,$ $m$, and $\nu $ satisfy the
relation $\sigma =2\left( m+\nu \right) $. In this case, we can state the
following result on the form of the coherent states $\left( 5.1\right)
.\medskip $

\textbf{Proposition 5.1.}\textit{\ Let }$2\nu >1$ \textit{and} $\sigma
=2\left( m+\nu \right) $\textit{.} \textit{Then, the wave functions of the
coherent states (5.1) are of the form}
\begin{equation*}
<\rho \mid z;m,\nu >=\frac{\sqrt{2\Gamma \left( 2\nu +m\right) }\left(
m!\right) ^{-\frac{1}{2}}}{\Gamma \left( \nu +\frac{1}{2}\right) \Gamma
\left( 2\nu \right) }\frac{\left( 1-z\overline{z}\right) ^{\nu }}{\left( 1-%
\overline{z}\right) ^{2\nu }}\left( \frac{z-1}{1-\overline{z}}\right)
^{m}\left( c^{2}\right) ^{-i\rho }\left( -\rho \right) ^{\left( \nu \right)
}\Gamma \left( \nu +i\rho \right)
\end{equation*}
\begin{equation}
\times F_{5}\left(
\begin{array}{ccc}
\nu +i\rho & \nu -i\rho : & 2\nu +m \\
. & \nu +\frac{1}{2}: & 2\nu
\end{array}
\mid \frac{1-\overline{z}z}{\left( 1-z\right) \left( \overline{z}-1\right) },%
\frac{1}{1-z}\right)  \tag{5.2}
\end{equation}
\textit{for any }$\rho \in \mathbb{R}_{+},$ \textit{where }$\digamma _{5}$%
\textit{\ is a special Appel-Kamp\'{e} de F\'{e}riet hypergeometric
function.\medskip }

\textbf{Proof. }We start from $\left( 5.1\right) $ by replacing the
coefficients $f_{k}^{(\sigma ,m)}\left( z\right) $ by their expressions in $%
\left( 2.10\right) .$ This leads to the expression
\begin{equation}
<\rho \mid z;m,\nu >=\left( \frac{(2\nu -1)}{\pi \left( 1-z\overline{z}%
\right) ^{\sigma }}\right) ^{-\frac{1}{2}}\sum_{k=0}^{\infty }\sqrt{\frac{%
\left( 2\nu -1\right) k!\Gamma \left( 2\nu +m\right) }{\pi m!\Gamma \left(
2\nu +k\right) }}  \tag{5.3}
\end{equation}
\begin{equation*}
\times \left( 1-z\overline{z}\right) ^{-m}\left( -1\right)
^{k}z^{m-k}P_{k}^{\left( m-k,2\nu -1\right) }\left( 1-2z\overline{z}\right)
\psi _{k}^{\nu }\left( \rho \right) .
\end{equation*}
Next, setting $u:=2z\overline{z}-1$ and inserting the expression $\left(
4.9\right) $ into $\left( 5.3\right) $, we obtain that
\begin{equation*}
<\rho \mid z;m,\nu >=\frac{\sqrt{\Gamma \left( 2\nu +m\right) }}{\sqrt{m!}}%
\left( 1-z\overline{z}\right) ^{\nu }z^{m}\sqrt{2}\left( c^{2}\right)
^{-i\rho }\left( -\rho \right) ^{\left( \nu \right) }\Gamma \left( \nu
+i\rho \right)
\end{equation*}
\begin{equation}
\times \sum_{k=0}^{\infty }\frac{z^{-k}}{\Gamma \left( 2\nu +k\right) \Gamma
\left( \nu +\frac{1}{2}+k\right) }P_{k}^{\left( 2\nu -1,m-k\right) }\left(
u\right) S_{k}\left( \rho ^{2},\gamma ,\gamma ,1/2\right) .  \tag{5.4}
\end{equation}
Next, we set $t:=1/z$ and denote the sum in $\left( 5.4\right) $ by
\begin{equation}
\frak{S:}=\sum_{k=0}^{\infty }\frac{t^{k}}{\Gamma \left( 2\nu +k\right)
\Gamma \left( \nu +\frac{1}{2}+k\right) }P_{k}^{\left( 2\nu -1,m-k\right)
}\left( u\right) S_{k}\left( \rho ^{2},\nu ,\nu ,1/2\right) .  \tag{5.5}
\end{equation}
By using (A26) of the Appendix, the sum $\left( 5.5\right) $ can be
expressed as
\begin{equation}
\frak{S}=\frac{\left( 1-\overline{z}\right) ^{-2\nu }z^{-m}}{\Gamma \left(
\nu +\frac{1}{2}\right) \Gamma \left( 2\nu \right) }\left( \frac{z-1}{1-%
\overline{z}}\right) ^{m}F_{5}\left(
\begin{array}{ccc}
\nu +i\rho & \nu -i\rho : & 2\nu +m \\
. & \nu +\frac{1}{2}: & 2\nu
\end{array}
\mid \frac{1-\overline{z}z}{\left( 1-z\right) \left( \overline{z}-1\right) },%
\frac{1}{1-z}\right)  \tag{5.6}
\end{equation}
and we therefore arrive at the expression $\left( 5.2\right) $ as stated in
the proposition. $\blacksquare \medskip $

\textbf{Corollary 5.1}\textit{. For }\textbf{\ }$m=0$\textit{, the wave
functions of the coherent states (5.1) are given in terms of the Gauss\
hypergeometric }$_{2}F_{1}$\emph{-sum by}\textit{\ }
\begin{equation}
<\rho \mid z;0,\nu >=\frac{\sqrt{2}\left( c^{2}\right) ^{-i\rho }\left(
-\rho \right) ^{\left( \nu \right) }\Gamma \left( \nu +i\rho \right) }{%
\Gamma \left( \nu +\frac{1}{2}\right) \sqrt{\Gamma \left( 2\nu \right) }}%
\left( \frac{1-z\overline{z}}{1-\overline{z}}\right) ^{\nu }\left( 1-%
\overline{z}\right) ^{i\rho }\ _{2}F_{1}\left(
\begin{array}{c}
\nu +i\rho ,\frac{1}{2}+i\rho \\
\nu +\frac{1}{2}
\end{array}
\mid \overline{z}\right)  \tag{5.7}
\end{equation}
\textit{for any }$\rho \in \mathbb{R}_{+}$.\medskip

\textbf{Proof. }When $m=0$, the expression $\left( 5.2\right) $ reduces to
\begin{equation*}
<\rho \mid z,\nu >=\frac{\sqrt{2}}{\Gamma \left( \nu +\frac{1}{2}\right)
\sqrt{\Gamma \left( 2\nu \right) }}\frac{\left( 1-z\overline{z}\right) ^{\nu
}}{\left( 1-\overline{z}\right) ^{2\nu }}\left( c^{2}\right) ^{-i\rho
}\left( -\rho \right) ^{\left( \nu \right) }\Gamma \left( \nu +i\rho \right)
\end{equation*}
\begin{equation}
\times \digamma _{5}\left(
\begin{array}{ccc}
\nu +i\rho & \nu -i\rho : & 2\nu \\
. & \nu +\frac{1}{2}: & 2\nu
\end{array}
\mid \frac{1-\overline{z}z}{\left( 1-z\right) \left( \overline{z}-1\right) },%
\frac{1}{1-z}\right) .  \tag{5.8}
\end{equation}
Next, we can apply the reduction identity ($\left[ 9\right] $, p.136)$:$%
\begin{equation}
F_{5}\left(
\begin{array}{ccc}
c, & d: & a \\
. & e: & a
\end{array}
\mid \chi ,\zeta \right) =_{2}F_{1}\left(
\begin{array}{c}
c,d \\
e
\end{array}
\mid \chi +\zeta \right).  \tag{5.9}
\end{equation}
to rewrite the $\digamma _{5}$-sum in $\left( 5.8\right) $ as
\begin{equation}
\digamma _{5}\left(
\begin{array}{ccc}
\nu +i\rho & \nu -i\rho : & 2\nu \\
. & \nu +\frac{1}{2}: & 2\nu
\end{array}
\mid \frac{1-\overline{z}z}{\left( 1-z\right) \left( \overline{z}-1\right) },%
\frac{1}{1-z}\right) =_{2}\digamma _{1}\left(
\begin{array}{c}
\nu +i\rho ,\nu -i\rho \\
\nu +\frac{1}{2}
\end{array}
\mid \frac{\overline{z}}{\overline{z}-1}\right).  \tag{5.10}
\end{equation}
Finally, with the help of the Pffaf transformation ($\left[ 5\right] $, p$.$%
68)$:$
\begin{equation}
_{2}F_{1}\left(
\begin{array}{c}
a,b \\
c
\end{array}
\mid x\right) =\left( 1-x\right) ^{-a}\ _{2}F_{1}\left(
\begin{array}{c}
a,c-b \\
c
\end{array}
\mid \frac{x}{x-1}\right)  \tag{5.11}
\end{equation}
\ we can write the $_{2}\digamma _{1}$-sum in right hand side of $\left(
5.10\right) $ as
\begin{equation}
\left( 1-\overline{z}\right) ^{\nu +i\rho }\ _{2}F_{1}\left(
\begin{array}{c}
\nu +i\rho ,\frac{1}{2}+i\rho \\
\nu +\frac{1}{2}
\end{array}
\mid \overline{z}\right) .  \tag{5.12}
\end{equation}
Returning back to $\left( 5.8\right) $ and inserting $\left( 5.12\right)$%
, we arrive at the expression $\left( 5.7\right) $. $\blacksquare \medskip $

Now, since we have obtained the expression of the wave functions $\left(
5.2\right) $, we now can apply the coherent states formalism $\left[ 10%
\right] $ to define the transform $\mathcal{B}_{rel,\nu }^{m}:L^{2}\left(
\mathbb{R}_{+}\right) \rightarrow \mathcal{A}_{m}^{2\left( \nu +m\right)
}\left( \mathbb{D}\right) $ by
\begin{equation}
\mathcal{B}_{rel,\nu }^{m}\left[ \phi \right] \left( z\right) :=\left(
\mathcal{N}_{2\left( \gamma +m\right) ,m}\left( z\right) \right) ^{\frac{1}{2%
}}\left\langle \phi \mid z;m,\nu \right\rangle _{L^{2}\left( \mathbb{R}%
_{+}\right) }.  \tag{5.13}
\end{equation}
Precisely, we state the following.\medskip

\textbf{Theorem 5.1.} \textit{The coherent state transform associated with
the wave functions }$\left( 5.2\right) $\textit{\ is the isometry }$\mathcal{%
B}_{rel,\nu }^{m}:L^{2}\left( \mathbb{R}_{+}\right) \rightarrow \mathcal{A}%
_{m}^{2\left( \nu +m\right) }\left( \mathbb{D}\right) $ \textit{defined by}

\begin{equation}
\mathcal{B}_{rel,\nu }^{m}\left[ \phi \right] \left( z\right) =\frac{\left(
\frac{2\nu -1}{\pi }\right) ^{\frac{1}{2}}\sqrt{2\Gamma \left( 2\nu
+m\right) }}{\Gamma \left( \nu +\frac{1}{2}\right) \Gamma \left( 2\nu
\right) \sqrt{m!}}\left( \frac{1}{1-z}\right) ^{2\nu }\left( \frac{\overline{%
z}-1}{\left( 1-z\right) \left( 1-z\overline{z}\right) }\right) ^{m}
\tag{5.14}
\end{equation}
\begin{equation*}
\times \int\limits_{0}^{+\infty }\digamma _{5}\left(
\begin{array}{ccc}
\nu -i\rho & \nu +i\rho : & 2\nu +m \\
. & \nu +\frac{1}{2}: & 2\nu
\end{array}
\mid \frac{1-\overline{z}z}{\left( 1-\overline{z}\right) \left( z-1\right) },%
\frac{1}{1-\overline{z}}\right) \left( c^{2}\right) ^{i\rho }\frac{\Gamma
^{2}\left( \nu -i\rho \right) }{i^{\nu }\Gamma \left( -i\rho \right) }\phi
\left( \rho \right) d\rho
\end{equation*}
\textit{for any} $z\in \mathbb{D}$.\medskip

\textbf{Definition 5.2.}\textit{\ With the notation}\textbf{\ }$\nu =(1+%
\sqrt{1+2c^{4}})/2$, \textit{where }$c$\textit{\ denotes the light velocity},%
\textit{\ the coherent state transform }$\left( 5.14\right) $\textit{\ will
be called a true polyanalytic relativistic Bargmann transform of order }$m.$%
\textit{\medskip }

To recover the analytic case, one just have to proceed by a direct
replacement $m=0$ in $\left( 5.14\right) $ using the identity $\left(
5.9\right) $.\medskip

\textbf{Corollary 5.2. }\textit{For} $m=0$, \textit{the coherent state
transform }$\left( 5.14\right) $\textit{\ becomes the isometry }$\mathcal{B}%
_{rel,\nu }^{0}$ \textit{mapping the Hilbert space }$L^{2}\left( \mathbb{R}%
_{+}\right) $\textit{\ onto the Bergman space }$\mathcal{A}^{2\nu }\left(
\mathbb{D}\right) $\textit{\ by}
\begin{equation}
\mathcal{B}_{rel,\nu }^{0}\left[ \phi \right] \left( z\right) =\frac{\sqrt{%
\frac{2\nu -1}{\pi }}\left( 1-z\right) ^{-2\nu }\sqrt{2}}{\Gamma \left( \nu +%
\frac{1}{2}\right) \sqrt{\Gamma \left( 2\nu \right) }}\int\limits_{0}^{+%
\infty }\ _{2}\digamma _{1}\left(
\begin{array}{c}
\nu -i\rho ,\nu +i\rho \\
\nu +\frac{1}{2}
\end{array}
\mid \frac{z}{z-1}\right) \left( c^{2}\right) ^{i\rho }\frac{\Gamma
^{2}\left( \nu -i\rho \right) }{i^{\nu }\Gamma \left( -i\rho \right) }\phi
\left( \rho \right) d\rho  \tag{5.15}
\end{equation}
\textit{for any} $z\in \mathbb{D}$.\medskip

\textbf{Remark 5.1.} Taking into account the restrictions we have made on
physical parameters such as $\alpha $ $=\nu \in \mathbb{R}$ and $\omega
_{0}=c^{-2}$, we now can assert that the coherent states $\left( 5.7\right) $
of the analytic case $m=0$ coincide with those obtained by the authors in ($%
\left[ 11\right] ,$p.317) if one takes their labeling complex number $\zeta =%
\overline{z}$. Furthermore, from the analysis $\left[ 11\right] $, it clearly
appears that our obtained results concerning the polyanalytic case $m\neq 0$  (in addition to other known quantities such as the overlap relation $\left(3.6\right)$) could be exploited as basic tools in the problem of deriving
the path integral representation for the transition amplitude (propagator)
between the constructed coherent states $\left\{ \mid z;m,\nu >\right\} $ in
$\left( 5.2\right) $.\medskip

\begin{center}
\textbf{Appendix}
\end{center}

We start by considering the sum
\begin{equation}
\frak{S}=\sum_{k=0}^{\infty }\frac{t^{k}}{\Gamma \left( 2\nu +k\right)
\Gamma \left( \nu +\frac{1}{2}+k\right) }P_{k}^{\left( 2\nu -1,m-k\right)
}\left( u\right) S_{k}\left( \rho ^{2},\nu ,\nu ,1/2\right) .  \tag{A1}
\end{equation}
From $\left( 4.5\right) $ the continuous dual Hahn polynomials in $\left(
A1\right) $ read
\begin{equation}
S_{k}\left( \rho ^{2},\nu ,\nu ,1/2\right) =\frac{\Gamma \left( 2\nu
+k\right) \Gamma \left( \nu +\frac{1}{2}+k\right) }{\Gamma \left( 2\nu
\right) \Gamma \left( \nu +\frac{1}{2}\right) }\ _{3}F_{2}\left(
\begin{array}{c}
-k,\nu +i\rho ,\nu -i\rho \\
2\nu ,\nu +\frac{1}{2}
\end{array}
\mid 1\right) .  \tag{A2}
\end{equation}
So that the sum $\left( A1\right) $ becomes
\begin{equation}
\frak{S}=\frac{1}{\Gamma \left( 2\nu \right) \Gamma \left( \nu +\frac{1}{2}%
\right) }\sum_{k=0}^{\infty }t^{k}P_{k}^{\left( 2\nu -1,m-k\right) }\left(
u\right) \ _{3}F_{2}\left(
\begin{array}{c}
-k,\nu +i\rho ,\nu -i\rho \\
2\nu ,\nu +\frac{1}{2}
\end{array}
\mid 1\right) .  \tag{A3}
\end{equation}
We now make use of the integral formula ($\left[ 22\right] $, p.314)$:$
\begin{equation}
\int_{0}^{y}x^{\alpha -1}\left( y-x\right) ^{\beta -1}\ _{2}F_{1}\left(
\begin{array}{c}
a,b \\
c
\end{array}
\mid \frac{x}{y}\right) dx=\frac{\Gamma \left( \alpha \right) \Gamma \left(
\beta \right) }{\Gamma \left( \alpha +\beta \right) }y^{\alpha +\beta -1}\
_{3}F_{2}\left(
\begin{array}{c}
a,b,\alpha \\
c,\alpha +\beta
\end{array}
\mid 1\right)  \tag{A4}
\end{equation}
with conditions $\left[ y\text{,}\Re \alpha >0,\Re \beta ,\Re \left(
c-a-b+\beta \right) >0\right] $. For parameters $y=1,$ $a=-k,b=\nu +i\rho
,\alpha =\nu -i\rho ,c=2\nu $ and $\beta =\frac{1}{2}+i\rho $ we see that $%
\left( A4\right) $ reads
\begin{equation}
_{3}F_{2}\left(
\begin{array}{c}
-k,\nu +i\rho ,\nu -i\rho \\
2\nu ,\nu +\frac{1}{2}
\end{array}
\mid 1\right) =\frac{\Gamma \left( \nu +\frac{1}{2}\right) }{\Gamma \left(
\nu -i\rho \right) \Gamma \left( \frac{1}{2}+i\rho \right) }%
\int\limits_{0}^{1}\frac{x^{-i\rho -1+\nu }}{\left( 1-x\right) ^{\frac{1}{2}%
-i\rho }}\ _{2}F_{1}\left(
\begin{array}{c}
-k,\nu +i\rho \\
2\nu
\end{array}
\mid x\right) dx.  \tag{A5}
\end{equation}
Therefore, $\left( A3\right) $ transforms as
\begin{equation}
\frak{S}=\frac{1}{\Gamma \left( 2\gamma \right) }\int\limits_{0}^{1}\frac{%
x^{\nu -i\rho -1}\left( 1-x\right) ^{-\frac{1}{2}+i\rho }}{\Gamma \left( \nu
-i\rho \right) \Gamma \left( \frac{1}{2}+i\rho \right) }\left(
\sum_{k=0}^{\infty }t^{k}P_{k}^{\left( 2\nu -1,m-k\right) }\left( u\right) \
_{2}F_{1}\left(
\begin{array}{c}
-k,\nu +i\rho \\
2\nu
\end{array}
\mid x\right) \right) dx.  \tag{A6}
\end{equation}
Now, we look closely at the sum in $\left( A6\right) $:
\begin{equation}
\Xi \left( x\right) :=\sum_{k=0}^{\infty }t^{k}P_{k}^{\left( 2\nu
-1,m-k\right) }\left( u\right) \ _{2}F_{1}\left(
\begin{array}{c}
-k,\nu +i\rho \\
2\nu
\end{array}
\mid x\right) .  \tag{A7}
\end{equation}
Using the connection formula ($\left[ 13\right] $, p$.555$)$:$%
\begin{equation}
P_{n}^{\left( \alpha ,\beta \right) }\left( u\right) =\left( \frac{1-u}{2}%
\right) ^{n}P_{n}^{\left( -2n-\alpha -\beta -1,\beta \right) }\left( \frac{%
u+3}{u-1}\right)  \tag{A8}
\end{equation}
for the parameters $\alpha =2\gamma -1,\beta =m-k$ and $n=k$ then $\left(
A8\right) $ takes the form
\begin{equation}
P_{k}^{\left( 2\gamma -1,m-k\right) }\left( u\right) =\left( 1-z\overline{z}%
\right) ^{k}P_{k}^{\left( -2\gamma -m-k,m-k\right) }\left( \frac{z\overline{z%
}+1}{z\overline{z}-1}\right) .  \tag{A9}
\end{equation}
Eq.$\left( A7\right) $ can then be rewritten as
\begin{equation}
\Xi =\sum_{k=0}^{\infty }\theta ^{k}P_{k}^{\left( -2\gamma -m-k,m-k\right)
}\left( V\right) \ _{2}F_{1}\left(
\begin{array}{c}
-k,\gamma +i\rho \\
2\gamma
\end{array}
\mid x\right) .  \tag{A10}
\end{equation}
in terms of the variables
\begin{equation}
\theta :=\frac{1-z\overline{z}}{z}\text{ and }V:=\frac{z\overline{z}+1}{z%
\overline{z}-1}.  \tag{A11}
\end{equation}
We now can apply the bilinear generating formula ($\left[ 23\right] $, p$.14$%
)$:$%
\begin{equation}
\sum_{k=0}^{\infty }\frac{\left( b\right) _{k}}{\left( d\right) _{k}}\theta
^{k}P_{k}^{\left( \alpha -k,\beta -k\right) }\left( V\right) \
_{2}F_{1}\left(
\begin{array}{c}
-k,c \\
b
\end{array}
\mid y\right)  \tag{A12}
\end{equation}
\begin{equation*}
=\left( 1-y\right) ^{-c}F_{8}\left( b,b,b,c,-\alpha ,-\beta ;b,d,d;\frac{y}{%
y-1},-\frac{1}{2}\left( V+1\right) \theta ,-\frac{1}{2}\left( V-1\right)
\theta \right) .
\end{equation*}
Note that the case $b=d=2\nu $ implies that the Lauricella triple
hypergeometric series $F_{8}$, which is denoted $F_{G}$ by in $\left[ 23%
\right] $, reduces to the expression
\begin{equation}
\left[ 1+\frac{\left( V+1\right) \theta }{2}\right] ^{\alpha }\left[ 1+\frac{%
\left( V-1\right) \theta }{2}\right] ^{\beta }\left( 1-y\right)
^{c}F_{1}\left( c,-\alpha ,-\beta ;b;\frac{y\left( V+1\right) \theta }{%
2+\left( V+1\right) \theta },\frac{y\left( V-1\right) \theta }{2+\left(
V-1\right) \theta }\right)  \tag{A13}
\end{equation}
in terms of the $F_{1}$ Appell's hypergeometric function ($\left[ 24\right] $%
, p.265)$.$ Therefore, in terms of our parameters, the sum $\left(
A10\right) $ has the following expression
\begin{equation}
\Xi =\frac{\left[ 1+\frac{\left( V-1\right) \theta }{2}\right] ^{m}}{\left[
1+\frac{\left( V+1\right) \theta }{2}\right] ^{2\nu +m}}\digamma _{1}\left(
\nu +i\rho ,2\nu +m,-m;2\nu ;\frac{x\left( V+1\right) \theta }{2+\left(
V+1\right) \theta },\frac{x\left( V-1\right) \theta }{2+\left( V-1\right)
\theta }\right).  \tag{A14}
\end{equation}
Denote the parameters and arguments occurring in the last $F_{1}$-sum
respectively by $a=\gamma +i\xi $, $b=2\gamma +m$, $c=-m$, $d=2\gamma $,
\begin{equation}
X=\frac{x\left( V+1\right) \theta }{2+\left( V+1\right) \theta }=x\mu _{z}%
\text{, \ \ \ \ \ \ \ \ \ \ \ \ \ \ }Y=\frac{x\left( V-1\right) \theta }{%
2+\left( V-1\right) \theta }=x\nu _{z}  \tag{A15}
\end{equation}
with
\begin{equation}
\mu _{z}:=\frac{\overline{z}}{\overline{z}-1}\text{, \ \ }\nu _{z}:=\frac{1}{%
1-z},  \tag{A16}
\end{equation}
then it can be presented here as $\digamma _{1}\left( a,b,c;d;X,Y\right) $
with the particularity that parameters $b,c$ and $d$ satisfy $d=b+c$. In
this case, we can apply the transformation ($\left[ 22\right] $, p.452):
\begin{equation}
\digamma _{1}\left( a,b,c,b+c;X,Y\right) =\left( 1-Y\right) ^{-a}\
_{2}\digamma _{1}\left(
\begin{array}{c}
a,b \\
b+c
\end{array}
\mid \frac{X-Y}{1-Y}\right) .  \tag{A17}
\end{equation}
Therefore, the $F_{1}$-sum in $\left( A14\right) $ reduces to
\begin{equation}
F_{1}\left( \nu +i\rho ,2\nu +m,-m;2\nu ;x\mu _{z},x\nu _{z}\right)
\tag{A18}
\end{equation}
\begin{equation*}
=\left( 1-x\nu _{z}\right) ^{-\left( \nu +i\rho \right) }\ _{2}F_{1}\left(
\begin{array}{c}
\nu +i\rho ,2\nu +m \\
2\nu
\end{array}
\mid \frac{(\mu _{z}-\nu _{z})x}{1-x\nu _{z}}\right) .
\end{equation*}
We set
\begin{equation}
\tau _{z}:=\mu _{z}-\nu _{z}=\frac{1-\overline{z}z}{\left( 1-z\right) \left(
\overline{z}-1\right) }.  \tag{A19}
\end{equation}
So that $\left( A14\right) $ becomes
\begin{equation}
\Xi \left( x\right) =\frac{\left[ 1+\frac{\left( V-1\right) \theta }{2}%
\right] ^{m}}{\left[ 1+\frac{\left( V+1\right) \theta }{2}\right] ^{2\nu +m}}%
\left( 1-x\nu _{z}\right) ^{-\nu -i\rho }\ _{2}F_{1}\left(
\begin{array}{c}
\nu +i\rho ,2\nu +m \\
2\nu
\end{array}
\mid \frac{x\tau _{z}}{1-\nu _{z}x}\right) .  \tag{A20}
\end{equation}
After computing the prefactor in the right hand side of $\left( A20\right) $
we find that

\begin{equation}
\Xi \left( x\right) =\left( \frac{z-1}{1-\overline{z}}\right) ^{m}\frac{%
z^{-m}\left( 1-\overline{z}\right) ^{-2\nu }}{\left( 1-x\nu _{z}\right)
^{\nu +i\rho }}\ _{2}F_{1}\left(
\begin{array}{c}
\nu +i\rho ,2\nu +m \\
2\nu
\end{array}
\mid \frac{x\tau _{z}}{1-x\nu _{z}}\right).  \tag{A21}
\end{equation}
By $\left( A6\right) $ and $\left( A20\right)$, it follows that
\begin{equation}
\frak{S}=\frac{\left( \frac{z-1}{1-\overline{z}}\right) ^{m}z^{-m}\left( 1-%
\overline{z}\right) ^{-2\nu }}{\Gamma \left( 2\nu \right) \Gamma \left( \nu
-i\rho \right) \Gamma \left( \frac{1}{2}+i\rho \right) }\int%
\limits_{0}^{1}x^{\nu -i\rho -1}\frac{\left( 1-x\right) ^{-\frac{1}{2}+i\rho
}}{\left( 1-x\nu _{z}\right) ^{\nu +i\rho }}\ _{2}F_{1}\left(
\begin{array}{c}
2\nu +m,\nu +i\rho \\
2\nu
\end{array}
\mid \frac{\tau _{z}x}{1-\nu _{z}x}\right) dx.  \tag{A22}
\end{equation}
A this stade we can make use of the integral representation ($\left[ 9\right]
$, p.137) of a very special case of the Appel-Kamp\'{e} de F\'{e}riet's
hypergeometric function $\left[ 25\right] $
as follows
\begin{equation}
F\left[
\begin{array}{ccc}
2 & c & d \\
1 & a & b \\
1 & e & . \\
1 & a^{\prime } & b
\end{array}
\mid
\begin{array}{c}
\chi \\
\zeta
\end{array}
\right] \equiv F_{5}\left(
\begin{array}{ccc}
c, & d: & a \\
. & e: & a^{\prime }
\end{array}
\mid \chi ,\zeta \right)  \tag{A23}
\end{equation}
\begin{equation}
=\frac{\Gamma \left( e\right) }{\Gamma \left( d\right) \Gamma \left(
e-d\right) }\int\limits_{0}^{1}t^{d-1}\left( 1-t\right) ^{e-d-1}\left(
1-\zeta t\right) ^{-c}\ _{2}F_{1}\left(
\begin{array}{c}
a,c \\
a^{\prime }
\end{array}
\mid \frac{\chi t}{1-\zeta t}\right) dt  \tag{A24}
\end{equation}
for parameters $d=\nu -i\rho $, $e=\nu +\frac{1}{2}$, $c=\nu +i\rho $, $%
a=2\nu +m$, $a^{\prime }=2\nu ,\zeta =\nu _{z}$, $\chi =\tau _{z}$ and $t=x.$
So that the integral occurring in $\left( A22\right) $ reads
\begin{equation}
\frac{\Gamma \left( \nu -i\rho \right) \Gamma \left( \frac{1}{2}+i\rho
\right) }{\Gamma \left( \nu +\frac{1}{2}\right) }F_{5}\left(
\begin{array}{ccc}
\nu +i\rho & \nu -i\rho : & 2\nu +m \\
. & \nu +\frac{1}{2}: & 2\nu
\end{array}
\mid \tau _{z},\nu _{z}\right) ,  \tag{A25}
\end{equation}
and therefore the sum $\frak{S}$ we are looking for is of the form
\begin{equation}
\frak{S}=\frac{\left( 1-\overline{z}\right) ^{-2\nu }z^{-m}}{\Gamma \left(
\nu +\frac{1}{2}\right) \Gamma \left( 2\nu \right) }\left( \frac{z-1}{1-%
\overline{z}}\right) ^{m}F_{5}\left(
\begin{array}{ccc}
\nu +i\rho & \nu -i\rho : & 2\nu +m \\
. & \nu +\frac{1}{2}: & 2\nu
\end{array}
\mid \frac{1-\overline{z}z}{\left( 1-z\right) \left( \overline{z}-1\right) },%
\frac{1}{1-z}\right) .  \tag{A26}
\end{equation}

\textbf{References}

{\footnotesize{$\left[ 1\right] $\ V. Bargmann, On a Hilbert space of analytic
functions and an associated integral transform, Part I. \textit{Comm. Pure.
Appl. Math}.14 (1961) 187}}

{\small $\left[ 2\right] $\ F. ELWassouli, A. Ghanmi, A. Intissar and Z.
Mouayn, Generalized second Bargmann transforms associated with the
hyperbolic Landau levels on the Poincar\'{e} disk, \textit{Ann. Henri
Poincar\'{e}.}\textbf{\ 13} (2012) 513 }

{\small $\left[ 3\right] $\ Mourad E.H. Ismail, Classical and Quantum
Orthogonal Polynomials in one variable, Encyclopedia of Mathematics and its
applications, Cambridge university press (2005) }

{\small $\left[ 4\right] $\ I. I. Goldman, I. I and D. V. Krivchenkov,
Problems in Quantum Mechanics. Pergamon, London (1961) }

{\small $\left[ 5\right] $\ G. E. Andrews, R. Askey and R. Roy, Special
Functions, Encyclopedia of Mathematics and its Applications, Cambridge
University Press, 1999 }

{\small $\left[ 6\right] $\ S. M. Nagiyev, E.I. Jafarov and R. M. Imanov,
The relativistic linear singular oscillator,\textit{\ J. Phys. A: Math. Gen}%
. \textbf{36} (2003) 7813 }

{\small $\left[ 7\right] $\ Z. Mouayn, Husimi's Q-function for the isotonic
oscillator in a generalized negative binomial states representation, \textit{%
Math. Phys. Anal. Geom.} 17 (2014) 289}

{\small $\left[ 8\right] $\ L. D. Abreu and H. G. Feichtinger, Functions
spaces of polyanalytic functions, Harmonic and Complex Analysis and its
Applications, Trends in Mathematics A. Vasilev (Ed.), Springer (2014) 139. }

{\small $\left[ 9\right] $\ S. k. Kulshreshtha, On Appell's double
hypergeometric functions, \textit{Collectanea Mathematica}. \textbf{19} (3)
(1968) 135}

{\small $\left[ 10\right] $\ J.P. Gazeau, Coherent states in quantum
physics, Wiley-VCH Verlag GmbH \& KGaA Weinheim, 2009 }

{\small $\left[ 11\right] $\ S. M. Nagiyev, E. I. Jafarov and M. Y.
Efendiyev, Coherent states and and a Path integral for the relativistic
linear singular oscillator, \textit{Commun. Theor. Phys.} \textbf{49} (2008) 315 }

{\small $\left[ 12\right] $\ Z. Mouayn, Coherent states attached to Landau
levels on the Poincar\'{e} disk, \textit{J. Phys. A: Math \& Gen}, \textbf{38%
} (2005) 9309}

{\small $\left[ 13\right] $\ Y. A. Brychkov, Handbook of Special Functions
Derivatives Integrals Series and Other Formulas, Taylor \& Francis 2008 }

{\small $\left[ 14\right] $\ N. Ikeda and H. Matsumoto, Brownian motion on
the hyperbolic plane and Selberg trace formula, \textit{J. Funct. Anal}.
\textbf{163  } (1999) 63 }

{\small $\left[ 15\right] $\ A. Comtet, On Landau levels on the hyperbolic
plane. \ \textit{Ann. Phys}. \textbf{173} (1987)\ 185}

{\small $\left[ 16\right] $\ S. M. Barnett,\ Negative binomial states of the
quantized radiation field,} {\small \textit{J. Mod. Opt. A}. \textbf{45 }%
(1998)\ 2201}

{\small $\left[ 17\right] $\ S. T. Ali, J. P. Antoine, J. P. Gazeau,
Coherent states and their generalizations, Springer, Berlin 2000}

{\small $\left[ 18\right] $\ J. R. Klauder, Continuous Representation theory
I. Postulates of continuous representation theory,\textit{\ J. Math. Phys}.\
\textbf{4} (1963) 1055}

{\small $\left[ 19\right] $\ Freeman M., Mateev M.D. and Mir-Kasimov R.M.,
\textit{Nucl. Phys. B}.\textbf{\ 12 }(1969) 197 }

{\small $\left[ 20\right] $\ S. M. Nagiyev, E. I. Jafarov and M. Y.
Efendiyev, Coherent states and a Path integral for the relativistic linear
singular oscillator, \textit{Commun. Theor. Phys}. \textbf{49} (2008) 315 }

{\small $\left[ 21\right] $\ S. M. Nagiyev, E. I. Jafarov and R. M. Imanov,
On a dynamical symmetry group of the relativistic linear singular
oscillator, \textit{Europhys. Lett.} \textbf{76} (2006) 175}

{\small $\left[ 22\right] $\ A. P. Prudnikov, Y. A. Brychkov and O.I.
Marichev, Integrals and Series, Vol.3: More special Functions, Gordon and
Breach, New York 1990 }

{\small $\left[ 23\right] $\ S.Saran, Theorems on bilinear generating
functions,\ \textit{Indian J. Pure. Appl. Math.} \textbf{3} (1972) 12}

{\small $\left[ 24\right] $\ E. D. Rainville, Special functions, The
Macmillan company, New York 1963 }

{\small $\left[ 25\right] $\ P. Appell and J. Kamp\'{e} de F\'{e}riet,
Fonctions hyperg\'{e}om\'{e}triques et hyperspheriques, Gauthier-Villars
1926 }

\end{document}